\documentclass[aps,prl,amssymb,amsmath,twocolumn,floatfix,pre]{revtex4}
\usepackage{graphicx}
\usepackage{amsmath}
\usepackage{color}

\begin{document}
\title{Comparison of the microcanonical population annealing algorithm with the Wang-Landau algorithm}
\author{Vyacheslav Mozolenko$^{1,2}$}
\author{Marina Fadeeva$^2$}
\author{Lev Shchur$^{1, 2}$}
\affiliation{$^1$ Landau Institute for Theoretical Physics, 142432 Chernogolovka, Russia}
\affiliation{$^2$ HSE University, 101000 Moscow, Russia}

\date{\today}

\begin{abstract}
The development of new algorithms for simulations in physics is as important as the development of new analytical methods. 
In this paper, we present a comparison of the recently developed microcanonical population annealing (MCPA) algorithm with the rather mature Wang-Landau algorithm.  The comparison is performed on two cases of the Potts model that exhibit a first-order phase transition. We compare the simulation results of both methods with exactly known results, including the finite-dimensional dependence of the maximum of the specific heat capacity. We evaluate the Binder cumulant minimum, the ratio of peaks in the energy distribution at the critical temperature, the energies of the ordered and disordered phases,  and interface tension. Both methods exhibit similar accuracy at selected sets of modeling parameters.  
\end{abstract}

\maketitle

\section{Introduction}
\label{sec:intro}

Numerical methods for investigating statistical mechanical models consist of a wide range approaches, among them~\footnote{The references cited in this paragraph reflect the authors' subjective view on the topic of modeling classical spin models and do not pretend to cover the entire wide range of the many good and informative papers.} low- and high-temperature series~\cite{Enting1,Butera2}, transfer-matrix methods~\cite{Blote3,Derrida4}, Markov Chain Monte Carlo (MCMC)~\cite{Landau-Binder-book,Geyer1992}, cluster Swendsen-Wang and Wolff methods~\cite{Swendsen-Wang,Wolff}, parallel tempering~\cite{SW-par,Earl-2005,Malakis-2013}, multicanonical methods (MUCA)~\cite{Berg-1993,Janke-1998}, Wang-Landau method~\cite{Wang-Landau,Wang-Landau2}, simulated annealing methods~\cite{Kirkpatrick-1983},  population annealing methods~\cite{Machta-2010,PA-Review}, etc. A recent systematic review of methods can be found in Ref.~\cite{Henin-2022}.
A common feature of these approaches is the explicit temperature dependence of the modeling process, with the exception of the Wang-Landau method. 
Temperature dependence leads to a critical slowing down of the simulation in the region around the critical point of the continuous transition~\cite{Landau-Binder-book}, which leads to a power law dependence of the correlation time on the system size, $t{\propto}L^z$, with the value of $z$ typically around $2$ for local MCMC~\cite{Wansleben} and an order of magnitude lower for cluster algorithms~\cite{Li-Sokal}. The practical computation time of a $d$-dimensional system, measured as the number of operations per spin, increases as $L^{d+z}$. Recent discussions of the dynamical critical exponent can be found in~\cite{Hasenbusch-2020,Groszek-2021,Liu-2023,Vatansever-2024}.
 
The modeling of discontinuous phase transitions in the critical region is determined by the interface between the regions of ordered and disordered phases. The surface energy of which is proportional to the length of the interface,  forming an energy barrier that depends exponentially on the length of the interface. This makes the simulation unpredictably slow in the critical region.
 Parallel tempering methods developed originally for spin glasses~\cite{SW-par} are unfortunately not effective in the case of first-order phase transitions.

The Wang-Landau algorithm has no temperature-dependent probabilities and uses the current estimate of the density of states (DoS) to calculate the probability of transition from one energy level to another. This algorithm relaxes the system to the true DoS~\cite{BFS}. The  relaxation process converges steadily with an appropriate modification~\cite{Bel-Per} of the original Wang-Landau algorithm and convergences as slowly as $1/t$~\cite{Liang,Liang-2}, with the time $t$ measured in elementary steps of a single-spin flip.

There is an interesting relation between microcanonical and multicanonical algorithms. DoS estimation is an essential part of the more complex MUCA~\cite{Berg-1993,Janke-1998} algorithm, and DoS estimation by the Wang-Landau algorithm can be incorporated into the MUCA scheme for weight estimation~\cite{Berg-2003}.

Another algorithm that has no temperature-dependent probabilities~\cite{Rose-2019} was recently presented, the microcanonical population algorithm. The special feature of this algorithm, is that it does not use any probabilities in the annealing part.  Instead, it uses the  {\em ceiling} of energy when moving down the energy spectrum or the {\em floor} of energy when moving up the energy spectrum~\cite{MS-2024}. The next feature is the parallel simulation of a huge number of replicas at a given energy, which allows us to estimate the energy dependence of thermodynamic functions. Together with the DoS estimation, this gives their temperature dependence.  We want to warn readers against a possible misunderstanding - in this note we discuss the microcanonical method of population annealing, in which energy is used as an annealing parameter. It should not be taken as an earlier and widely studied method of population annealing~\cite{Machta-2010,PA-Review}, in which temperature is used as an annealing parameter.

It seems that there is no critical slowdown in the usual sense in microcanonical algorithms because of the absence of an explicit temperature dependence of the algorithm on the temperature.  In the case of the first-order phase transition, the condensation and evaporation of droplets in the vicinity of the critical temperature are still determined by the energy barrier associated with surface tension and depend exponentially on the surface length. Microcanonical simulations are probably less sensitive to this than canonical simulations~\cite{Schierz-2016,Rose-2019}. There are examples of comparison of energy barriers in the Lennard-Jones particle system showing the same scaling behavior, but the barrier height is almost six times lower for the microcanonical ensemble compared to the canonical one~\cite{Janke-2018}.

In this paper, a comparative study of two algorithms is carried out, the modified Wang-Landau algorithm (WL-1/t)~\cite{Wang-Landau,Bel-Per} and the multicanonical population algorithm (MCPA)~\cite{Rose-2019,MS-2024}, is carried out. The reason for the comparative study is to understand the accuracy of the methods. 
As an example, we simulate the Potts model with 10 and 20 components that undergoes a strong discontinuous transition. We use the elementary step of a single-spin flip as the Monte Carlo step. The elementary step is very different in the two models due to the intrinsic difference of the models. 
We model a replica of the system with the WL-1/t algorithm and $R$ parallel replicas of the system with the MCPA algorithm. So, such defined Monte Carlo steps are more or less proportional to the simulation time. The details are discussed in the Conclusions Section of the paper.

\section{Model}
\label{sec:model} 
The Potts model~\cite{Potts,Wu-review} with $q$ state is a generalization of the Ising model consisting of N interacting spins ${\sigma_i\; (i{=} 1,...,N)}$, each of which takes values $\sigma_i {\in} {(1, ...,q})$. The energy $E$ is given by
\begin{equation}
    E = -J\sum_{\langle i,j \rangle} \delta(\sigma_i, \sigma_j)
    \label{eq:H}
\end{equation}
and for simplicity we set the ferromagnetic constant $J$ equal to unity $J{=}1$.

The critical temperature is known analytically (see~\cite{Baxter-book})
\begin{equation}
    \beta_c=\frac{J}{k_BT_c}=\ln(1+\sqrt{q}).
    \label{eq:t_crit}
\end{equation}

For $q{>}4$, the Potts model undergoes a first order phase transition and a mixing of ordered and disordered phases is observed between the respective energies $e_o$ and $e_d$.  In the thermodynamic limit $L{\rightarrow}\infty$ these energies are also known (see~\cite{Baxter-book})

\begin{equation}
    \frac{e_o+e_d}2=-\left( 1+\frac{1}{\sqrt{q}}\right)
    \label{eq:e_mean}
\end{equation}
and
\begin{equation}
    e_d - e_o = 2\left( 1+\frac{1}{\sqrt{q}}\right)\tanh\left( \frac{\Theta}{2}\right) \prod^{\infty}_{n=1}{\tanh^2(n\Theta)},
    \label{eq:e_delta}
\end{equation}
where $\Theta$ is defined as $2\cosh(\Theta) {=} \sqrt{q}$. 

\section{Direct estimation of DoS}
\label{sec-direct}

Procedures of calculating thermodynamic quantities from the density of states (DoS) are based on the representation of a partition function, replacing the summation over all possible spin configurations $\{j\}$
\begin{equation}
Z = \sum_{\{j\}} e^{-E(\{j\})/ k_B T},
\end{equation}
where $k_B$ is the Boltzmann constant, $T$ is the temperature, with the summation over the 
 energy levels $n$
\begin{equation}
Z = \sum_{\{n\}} g(E_n)e^{-E_n/ k_B T}.
\label{Z-part}
\end{equation}
The function $g(E)$ is the density of states (DoS), that is, $g(E_n)$ is the number of configurations with energy $E_n$. 

There are two algorithms for direct estimation of DoS (technically the logarithm of DoS), the Wang-Landau algorithm~\cite{Wang-Landau,Wang-Landau2} and the microcanonical population annealing algorithm~\cite{Rose-2019,MS-2024}, which are the subject of a comparative study in this paper.

\subsection{Wang-Landau algorithm}
\label{sec:WL}

The Wang-Landau algorithm~\cite{Wang-Landau,Wang-Landau2}  is applicable to any system with a given partition function, which can be written as a sum over $n$ energy levels, as in Expr.~(\ref{Z-part}). This representation is key to the Wang-Landau algorithm.
Knowing $g(E)$, one can calculate, for example, the internal energy $E(\beta)$ and the heat capacity $C(\beta) $ at any value of the inverse temperature $\beta{=}1/k_BT$.
 
\begin{eqnarray}
  E(\beta) &= \langle E\rangle = \frac{\sum_{n=0}^{N_E-1} E_n g(E_n) e^{-\beta E_n}} {\sum_{n=0}^{N_E} g(E_n) e^{-\beta E_n}}, \label{eq-e}  \\     C(\beta) &= \beta^2 (\langle E^2 \rangle -  \langle E \rangle^2).   
    \label{eq-b}
\end{eqnarray} 

The basic idea of the algorithm is to organize a random walk through the energy space of the system using an appropriate transition probability for a random walk jump from the level energy $k$ to the energy level $m$, this Wang-Landau probability defined below in expr.~(\ref{PWL-expr}).

During the random walk~\cite{Wang-Landau},  two histograms $H(E)$ and $S(E)$ are accumulated.  The first histogram $S(E) {=} \log(g(E))$ contains the current value of the DoS logarithm. The second is the auxiliary histogram $H(E)$, which contains information about the number of visits to each energy level. At the beginning of the algorithm, $H(E_n)$ is initialized with zeros and $S(E_n)$ with ones. The initial configuration of the system is set to the ground state. The initial value of the modification parameter $f {=} f_0 {=} \exp(1) {\simeq} 2.71828$. 

The further steps of the algorithm are as follows: 1) the state of the system is changed (for the spin model it is a flip of the random spin) and the energy of the new state $E_m$ is calculated; 2) the transition from the state with energy $E_k$ to the state with energy $E_m$ takes place with the Wang-Landau probability
\begin{equation}
P_{WL}(E_k,E_m)=\min\left(1,\frac{\tilde S(E_k)}{\tilde S(E_m)}\right),
\label{PWL-expr}
\end{equation}
where $\tilde S(E_k) {=} \log(\tilde g(E_k))$ is the logarithm of the current DoS estimate. 
The next step is to update the auxiliary histogram $H(E_k) {\to} H(E_k) {+} 1$, and the current estimate $\tilde S(E_k) {\to} \tilde S(E_k){+}\ln(f)$. 

Steps 1) and 2) are repeated until the histogram $H(E_k)$ is sufficiently ``flat'', e.g., at the 5\%  level~\cite{Wang-Landau}). The value of the parameter $f$ is then updated as a function of the square root $f_i {=} \sqrt{f_{i-1}}$, and the histogram is reset $H(E_k) {=} 0$. Steps 1) and 2) are then repeated again. The algorithm ends when the parameter $f$ reaches some desired value of $f_{end}$, e.g., $f_{fin} {=} \exp(10^{-8}) {\simeq} 1.00000001$, as proposed in~\cite{Wang-Landau}.

\subsubsection{1/t-Wang-Landau algorithm}

The Wang-Landau algorithm is used in various fields of science, for example, in modeling polymers~\cite{polymer1,polymer2} and protein chains~\cite{proteinDNA,protein2}, for optimization~{\cite{Liang}. However, it is known that as the number of steps increases, the computational accuracy saturates at some step.  This has been observed in a number of publications such as~\cite{lee-landau,zhou-bhatt}. Consequently, it affects the accuracy of the calculation of thermodynamic functions for relatively large systems, including those in the critical region.  In addition, the choice of $\sqrt{f_i}$ as the function to change the parameter $f$ was left open until proposed in~\cite{Bel-Per} modification. 

 The 1/t-Wang-Landau algorithm is a modification~\cite{Bel-Per} that emerged as a solution to these problems.
  The first step of 1/t-Wang-Landau is similar to the original~\cite{Wang-Landau} method, except that criterion of histogram $H(E)$ ``flatness'' was replaced by the $H(E) {\ne} 0$ test. It is argued in~\cite{Bel-Per} that practically the flatness criteria can be replaced by a weaker but still global property, which is the requirement to visit each energy level before update the parameter $f$.
  
  The algorithm proceeds to the second stage when the condition $F_i {>} N_E/t$ is satisfied, where $F {=} \ln f$. From this point on, the parameter $F$ changes not as $F_i {=} F_{i-1}/2$  but $F_i {=} N_E/t$. Here, $t$ is the number of elementary spin flips and $N_E$ is the number of levels in the system. In the second stage, the histogram $H(E)$ is no longer checked. Theoretical justification for the convergence of this method was presented in the article~\cite{Liang}.
 
\subsubsection{Criterion of convergence}

In the original Wang-Landau algorithm~\cite{Wang-Landau} and its modification 1/t-Wang-Landau algorithm~\cite{Bel-Per} no method was proposed to evaluate the accuracy and convergence of the algorithm. The only argument for the applicability of the method is to demonstrate the quality of the data by comparing them with the exact solution known for the 2d Ising model~\cite{Beale-1996}.  A possible solution to this problem was proposed in~\cite{BFS}, where a square matrix $T(E_k, E_m)$ was additionally introduced instead of the histogram $H(E)$, and it was shown that the proximity of the matrix $T$ to stochastic can be used as an accuracy criterion for DoS estimation.
The matrix elements are the transition frequency from configurations with energy $E_k$ to configurations with energy $E_m$. The closer the estimated DoS is to the true DoS, the closer the matrix $T$ is to the stochastic one, and the deviation of the largest eigenvalue of the matrix from unity can be used as a criterion for the accuracy of the DoS estimate. Interestingly, this property of the matrix $T$ is consistent with the flatness criteria of the histogram $H(E)$ proposed by Wang and Landau and with the non-zero visitation criterion of each energy level proposed by Belardinelli and Pereyra. Thus, the matrix $T$ is more informative and, in addition, provides a clue for evaluating the accuracy and controlling the convergence of DoS.

\begin{figure}[h!]
    \centering
    \includegraphics[scale=0.25]{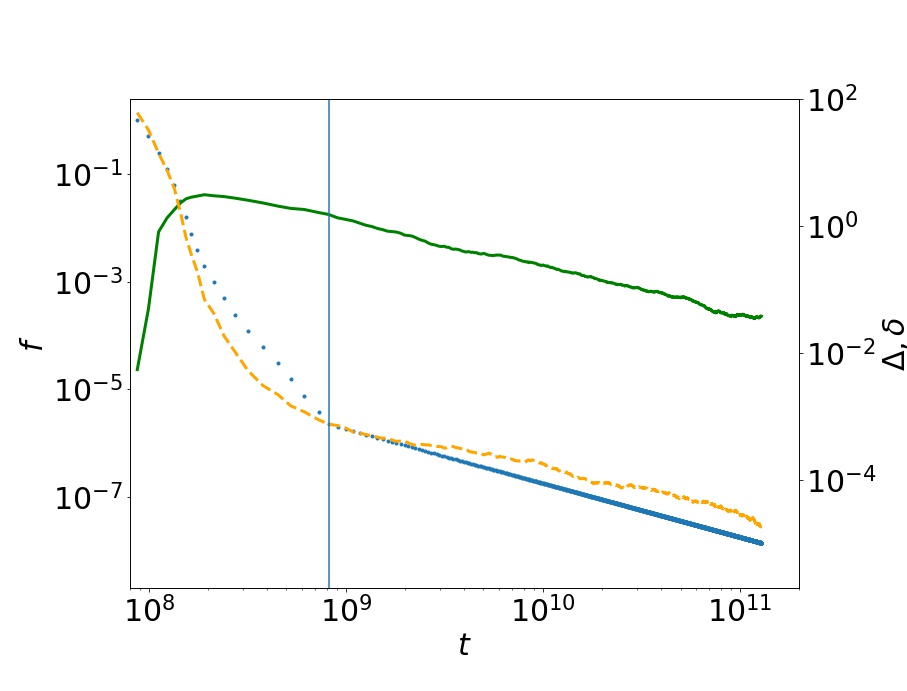} \includegraphics[scale=0.25]{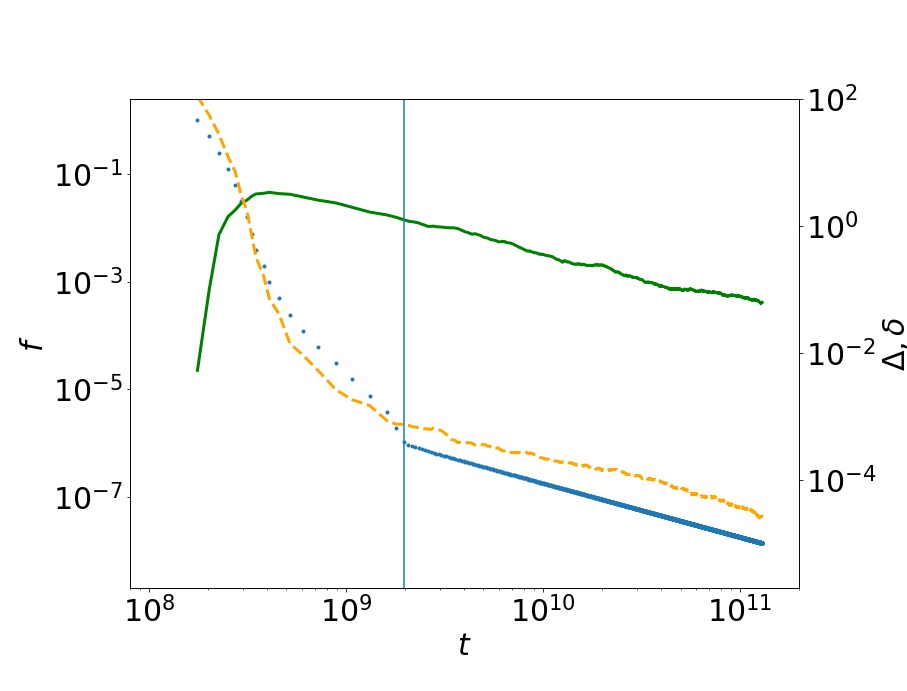}
        \caption{Dependence of the algorithm parameter $f$(dotted blue line and left axis),  convergence criteria $\delta$ (solid green line and right axis) and  $\Tilde{\Delta}$ (dashed yellow line and right axis) on the Monte-Carlo step $t$ for 10 state Potts model (top panel) and 20 state Potts model (bottom panel). Linear lattice size $L=30$. The values of $f$, $\delta$ and $\Tilde{\Delta}$ are the average over 20 runs. The vertical line is the average time $t_0$ for which the algorithm switches the function of the modification parameter $f$.}
    \label{fig1}
\end{figure}

The figures~\ref{fig1} show the variation of the algorithm parameter $f$, which is varied according to the Wang-Landau~\cite{Wang-Landau} rule $f_i{=}\sqrt{f_{i-1}}$ until the Belardinelli and Pereyra~\cite{Bel-Per}  criteria is met and then by the  $f{=}f_0/t$ rule. 
Also shown in the figures~\ref{fig1} is  the difference $\Delta$, which is calculated as the relative difference of DoS at simulation time $t$, $g(E_n;t)$ and at the final time $t_f$, $g(E_n;t_f)$
\begin{equation}
\Delta=\frac{1}{N_E}\sum_{n=1}^{N_E} \left( \frac{g(E_n;t)}{g(E_n;t_f)} - 1\right)
\label{eq-Delta}
\end{equation}
and the  convergence criterion $\delta$~\cite{BFS}
\begin{equation}
\delta=|1-\lambda_1|,
\label{eq-delta}
\end{equation}
where $\lambda_1$ is the largest eigenvalue of the transition matrix $T$, and the two simulation sets presented in Figure~\ref{fig1} refer to the 10 and 20 components of the Potts model. 

The  parameter $\Delta$ is well defined for models for which we have an exact solution for DoS, such as the Ising model in one and two dimensions. In our case with an exact DoS that is not available, we replace the unknown exact DoS values with the final DoS values $g(E_n;t_f)$ from the simulations. Time is measured in units of single-spin flip events. It can be seen that $\Delta$ decreases inversely with time, as shown by Liang~\cite{Liang}, and the control parameter $\delta$ decreases in the same way~\cite{BFS}. It should be noted that the $\delta$ convergence criterion can be used without knowledge of the DoS function and at any time step in the simulation. 

The $1/t$ dependence appears to be the optimal protocol for changing the algorithm parameter $f$~\cite{Bel-Per,Liang-2,BFS} in the last simulation steps. The fast convergence of the DoS in the initial simulation steps to the neighborhood of the exact DoS is still not understood, although the Wang-Landau  algorithm is widely used.

\subsubsection{Calculation with modified WL-1/t algorithm}

We use the Wang-Landau algorithm~\cite{Wang-Landau} with 1/t-modification~\cite{Bel-Per} and compute the convergence criterion $\delta$~\cite{BFS} during the simulation.

We simplify the protocol by fixing an interval of $M$ between the checks of the Bellardineli and Pereyra $H(E) {!=} 0$ criterion with $M{=}10^5$, which avoids multiple and unnecessary computations.  We also use this point to compute the largest eigenvalue of $T(E_k, E_m)$ using the procedure {\tt dgeev()}  from the Intel oneAPI Math Kernel Library LAPACK~\cite{lapacke}.  The random number generator {\tt mt19937.c}~\cite{mt,mt2} was used to generate a new configuration state and decide whether to accept the new configuration.

\subsection{Equilibrium microcanonical annealing algorithm}
\label{sec:MCPA}

A promising framework for simulating equilibrium systems in a microcanonical ensemble using annealing in an energy ceiling has been proposed by Rose and Machta and successfully applied to the first-order thermal transition in a 20-component two-dimensional Potts model with demonstration of topological transitions in the phase coexistence region~\cite{Rose-2019}.

\subsubsection{Rose-Machta ceiling procedure}

Here, we briefly introduce the Rose and Machta ceiling population algorithm presented in Section II of the paper~\cite{Rose-2019}. 

In Rose and Machta's approach to simulating equilibrium systems in a microcannonical ensemble, there is no relaxation on  temperature decrease; instead, the independent variable of the algorithm is energy. The MCMC procedure consists of a single spin-flip. Moves occur in configuration space, changing a randomly chosen spin, and the transition probability from an $\alpha$ state with energy $E_\alpha$ to a $\omega$ state with energy $E_\omega$ is defined as follows 
\begin{equation}
    P_{ceiling}(\alpha\rightarrow \omega) = \left\{
    \begin{array}{lcl}
        1 & {\rm if} & E_\omega \le E_c  \\ 
        0 & {\rm if} & E_\omega > E_c
    \end{array}
    \right.,
    \label{eq:prob-cooling}
\end{equation}
where $E_c$ is the ceiling energy value, the {\em cooling} energy value. An elementary MCMC step consists of updating $N$ of randomly chosen spins, where $N$ is the number of spins in the system. The number of elementary MCMC steps $n_s(E)$ is a parameter of the algorithm. 

The method is in a sense a mixture of three algorithms: the simulated annealing algorithm~\cite{Kirkpatrick-1983},  the Wang-Landau algorithm~\cite{Wang-Landau,Wang-Landau2}, and the population annealing algorithm~\cite{Machta-2010,PA-Review}.

\subsubsection{Floor microcanonical procedure}

A practical modification of the Rose-Machta algorithm has been augmented the floor energy~\cite{MS-2024} in addition to the ceiling energy, allowing the DoS to be estimated for all energy levels of the system. Here we basically repeat the relevant part of the paper~\cite{MS-2024}. The idea is to generate random configurations of the system that are most likely to correspond to energy levels in the neighborhood of the DoS maximum, which is a convex function. By applying the ceiling energy algorithm,  the left wing of the DoS can be estimated as the ceiling algorithm guides the replicas to the ground state. An extension of the ceiling algorithm, the floor algorithm~\cite{MS-2024}, instead directs replicas to a higher energy. 

In the floor algorithm, the probability of transition from the $\alpha$ state with energy $E_\alpha$ to the $\omega$ state with energy $E_\omega$ is defined as follows \begin{equation}
    P_{floor}(\alpha\rightarrow \omega) = \left\{
    \begin{array}{lcl}
        1 & {\rm if} & E_\omega \ge E_f  \\ 
        0 & {\rm if} & E_\omega < E_f
    \end{array}
    \right.,
    \label{eq:prob-heating}
    \end{equation}
where $E_f$ is the floor energy value.

\begin{figure}[h!]
    \centering
    \includegraphics[scale=0.25]{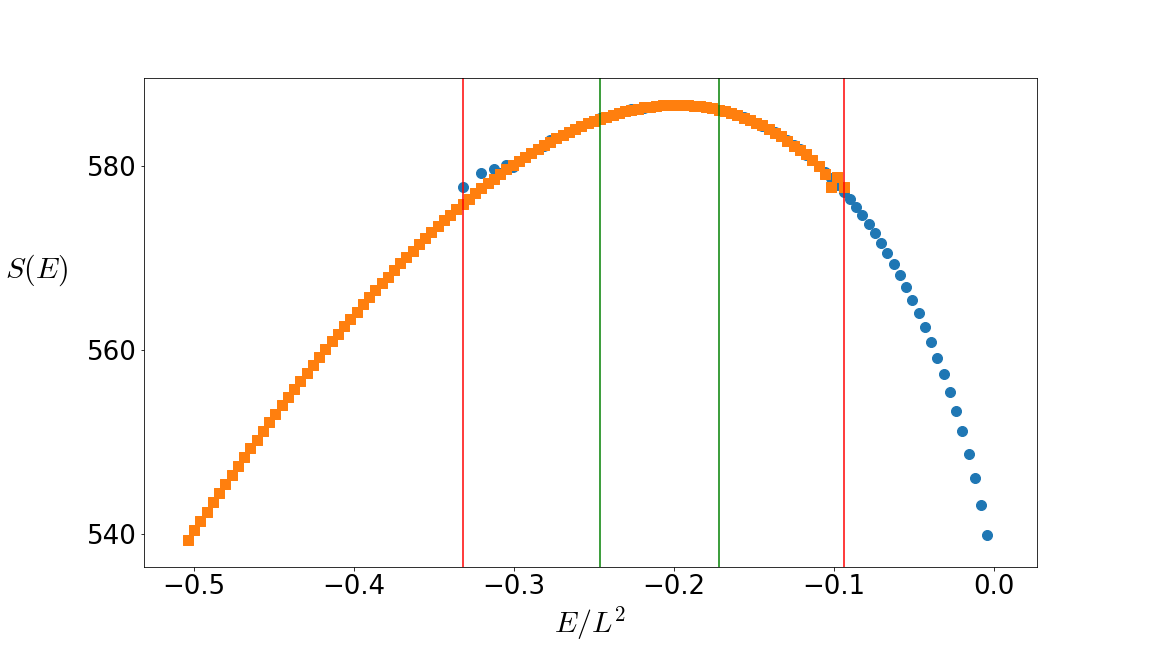}
    \caption{Illustration of the entropy estimation stitching procedure for the 10-component Potts model on a square lattice with linear size $L{=}16$. The orange squares correspond to the ceiling process and the blue circles to the floor process. The meaning of the vertical lines discussed in the text. }
    \label{fig2}
\end{figure}

\subsubsection{Microcanonical population annealing algorithm}

Combining the ceiling and floor procedures, we get the following algorithm. 

First we need to prepare a {\em pool} of replicas, for which we generate $R$ replicas of the system. The random distribution of spins in each replica will case the replica energies to have a value close to the maximum of the energy probability distribution $g(E)$, called the density of states (DOS). The first value of the ceiling $E_c$ will be the highest energy in the replica pool, and the first value of the floor $E_f$ will be the lowest energy in the replica pool.

In general, we do not know the energy spectrum in advance, and the next energy ceiling $E_c(i{+}1)$ or floor value $E_f(i{+}1)$  is chosen as the closest to the current energy from the energies of the current replica pool. Thus, we compute the energy spectrum of the system dynamically. 

The elementary step $i$ of the algorithm is as follows~\cite{MS-2024}

1. Select the new ceiling (floor) value. This could be the energy level down $E_c(i)$ from the current ceiling procedure or the energy level up $E_f(i)$ from the current floor procedure. 

2. Perform the MCMC check $n_s(E_c(i))$ or $n_s(E_f(i))$, thereby creating new configurations $R$ that represent a pool of configurations.

3. Count the number of replica in the pool $R'$ with energy $E_c(i)$ or $E_f(i)$, and calculate the culling fraction  $\epsilon(E_c(i)){=}R'/R$ or  $\epsilon(E_f(i)){=}R'/R$. Filter these $R'$ configurations from the pool of configurations. 

4. Randomly select with repetitions new replicas $R$ from the filtered pool of configurations.

The process was terminated if the condition $R'{=}R$ was satisfied, which means that the annealing reached the ground state in the ceiling protocol or the most symmetric energy level in the floor protocol. 

\subsubsection{Stitching entropy parts together}

To estimate the extensive part of entropy~\cite{Rose-2019}, the culling fractions  for the ceiling and floor are used
\begin{eqnarray}
S^c(E) &=& \ln(\epsilon(E)) + \sum_{E' > E} \ln(1 - \epsilon(E')), \\
S^f(E) &=& \ln(\epsilon(E)) + \sum_{E' < E} \ln(1 - \epsilon(E')).
\label{eq:S_calc}
\end{eqnarray}
Entropy allows us to add arbitrary constants, which we denote as $S_0^c$ and $S_0^f$, the entropy constants for the ceiling and floor, respectively.

As can be observed in a model example in figure~\ref{fig2}, both the cooling and heating runs cover only one wing of the whole spectrum. The intersection, covered by both runs, is placed near the maximum of entropy, where the random replica normally reside. We obtain entropy in the full energy range by stitching cooling and heating wings in the intersecting region.

Stitching is an arbitrary procedure and is not sensitive to the exact choice of it. We conduct it as follows:

1. Select the intersection area bounded by the red lines in Fig.~\ref{fig2} from the leftmost point of the ``heating'' wing to the rightmost point of the ``cooling'' wing.

2. The ends of the cooling and heating wings are somewhat scattered, so we cut off the outer parts of the region, leaving the region bounded by the green lines labeled $E_{left}$ and $E_{right}$. In this paper, we use a width between green lines three times smaller than between red lines, although this ratio can be changed depending on the specific task.

3. Calculate the average $\Delta S {=} avg(S^c(E) {-} S^h(E))$ for all energies in the last ``green'' area. This allows us to write cross-linked $S(E)$ in the form
\begin{equation}
    S(E) =  S_0 + \left\{
    \begin{array}{lcl}
        S^c(E)                             & {\rm if} \; E < E_{left} \\ 
        S^h(E) + \Delta S                  & {\rm if} \; E > E_{right} \\
        ( S^c(E) + S^h(E) + \Delta S ) / 2 & {\rm else}
    \end{array}
    \right.
    \label{eq:S_diff}
\end{equation}

4. The last free constant $S_0$, if required, can be established by counting the number of all states in the system, which in the case of the q-state Potts model is $q^{L^2}$, and $S_0$ should be chosen from the relation
\begin{equation}
    \sum_E e^{S(E)} = q^{L^2}.
    \label{eq:S_0}
\end{equation}

\subsubsection{Estimation of the thermodynamic observables}

An estimate of the partition function is given as a function of temperature $T$ measured in the energy units 
\begin{equation}
Z(T) = \sum_E  e^{- E/T  + S(E)}.
\label{eq:Z}
\end{equation}

The estimates of the average internal energy $\langle E(T)\rangle$, specific heat $\langle C(T)\rangle$, Binder cimulant~\cite{Binder-C}, and probability of energy distribution of $P(E;T)$ at temperature $T$ are calculated using the following expressions
\begin{eqnarray}
\langle E(T)\rangle  &=& \frac{\sum_E E\; e^{- E/T  + S(E)}}{Z(T)}, \label{eq:E} \\
\langle E^2(T)\rangle  &=& \frac{\sum_E E^2\; e^{- E/T  + S(E)}}{Z(T)}, \label{eq:E2}\\
C(T) &=& \frac{\langle E^2(T)\rangle - \langle E(T)\rangle^2}{T^2}, \label{eq:C}\\
V(T) &=& 1-\frac{\langle E^4(T)\rangle}{3\langle E^2(T)\rangle^2} \\
P(E;T) &=& \frac{ e^{- E/T  + S(E)}}{Z(T)}.
\label{eq:PET}
\end{eqnarray}

\subsection{Choice of the parameters}

Only two parameters can be chosen in the MCPA algorithm, the number of replicas $R$ and the number of Monte Carlo steps $n_s$. To demonstrate how these parameters affect the DoS accuracy, we simulate a two-dimensional Ising model for which the exact DoS $g^{exact}(E_n)$ is available~\cite{Beale-1996}, and compute the mean deviation $\delta g$

\begin{equation}
\delta g=\frac{1}{N_E}\sum_{n=1}^{N_E} \left( \frac{g(E_n;t_f)}{g^{exact}(E_n)} - 1\right).
\label{eq-Delta}
\end{equation}

We ran simulations for $R$ varying from $2^{10}$ to $2^{17}$, and for $n_s{=}1,5,10$ and 50. The results are shown in the figure~\ref{fig3} by symbols. The dependence on $R$ looks like $1/R^{1/2}$, which supports the idea that in the limit of sufficiently large $R$ the algorithm is ergodic near the ground state~\cite{Rose-2019} and works correctly, at least for the Ising model. 

\begin{figure}[h!]
    \centering
    \includegraphics[scale=0.25]{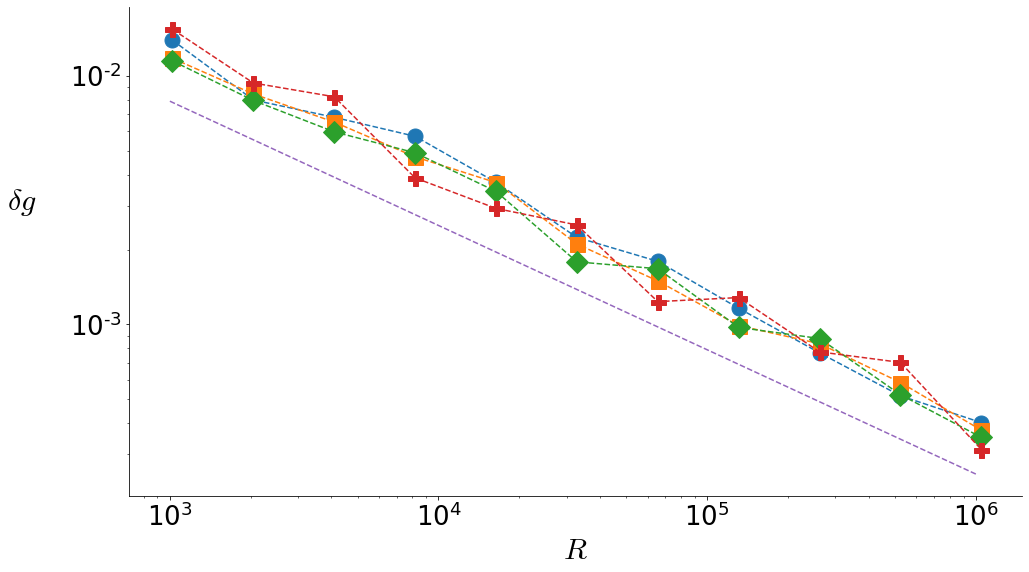}
    \caption{Accuracy illustration for an exactly solvable two-dimensional Ising model on a square lattice with linear dimension $L{=}20$, average deviation $\delta g$ of DoS. Blue circles correspond to $n_s = 1$, orange squares to $n_s=5$, green rhombuses to $n_s=10$, and red crosses to $n_s=50$. The dotted line corresponds to the slope $R^{-1/2}$.  }
    \label{fig3}
\end{figure}

\begin{figure}[h!]
    \centering
    \includegraphics[scale=0.3]{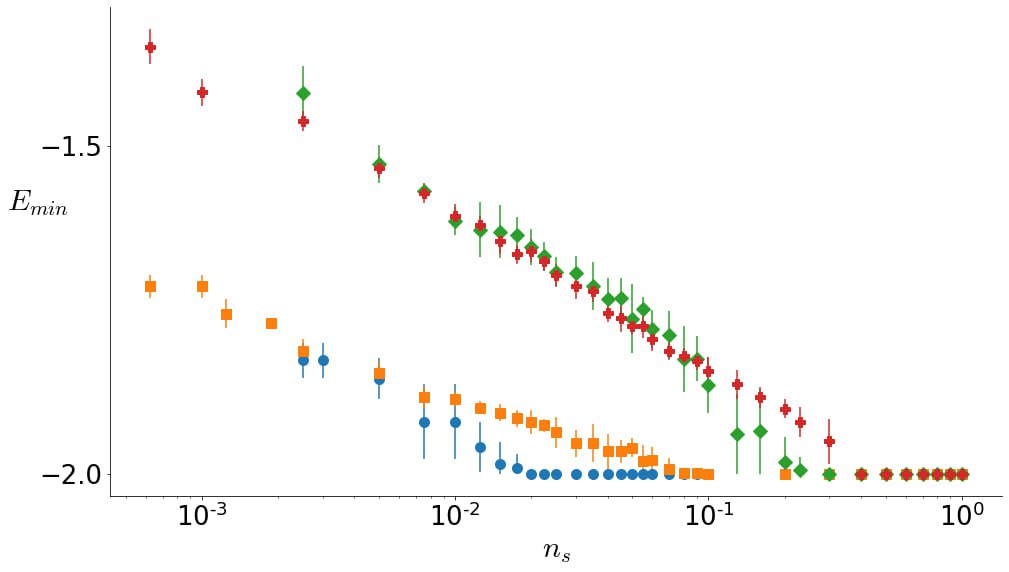}
    \caption{Illustration of the lowest energy level achieved. Dots represent the average over 10 runs, error bars represent the standard error. Blue circles represent the Ising model with $L {=} 20$, orange squares represent the Ising model with $L{=}40$, green diamonds represent the 10-q Potts model with $L{=}20$, and red crosses represent the 10-state Potts model with $L{=}40$.}
    \label{fig4}
\end{figure}

Depending on $n_s$, we sometimes can find ourself out of replicas before we reach ground state. 
Figure~\ref{fig4} shows how variation of $n_s$ influence the termination of the simulations with the lowest energy. With $n_s{=}1$ simulation with actually used number of replicas $R=2^{17}$ always reach ground state energy $e{=}{-}2$ for Ising model and 10 state Potts model (the same is valid for 20 state Potts model, not shown in the figure. 

As we see on a Fig.\ref{fig4}, we reach stable runs even before $n_s {=} 1$, and in this work we use $n_s{=}10$.
Anyway, we perform test simulations with the number of $n_s$ smaller than unity, which corresponds to the $L^2$ local spin flip attempts. These means that we use $n_s{<}L^2$, and one can see the linear convergence of simulation to the ground state as $n_s$ goes to 1. To be safe, we use in the following simulations $n_s{=}10$ with $R=2^{17}$.

\subsubsection{Calculation with MCPA algorithm}

The implementation of the MCPA algorithm is based on a modification~\cite{Rose-2019} of the accelerated population annealing algorithm for GPU~\cite{GPU-PA-2017}, presented in \cite{MS-2024}. The simulations were performed on an NVIDIA V100 GPU with a typical replica number $R{=}2^{17}{=}131072$. Spins are represented by a single number of type C 'char'.

We use the cuRAND~\cite{cuRAND} package with the Philox random number generator from the CUDA SL package, which allows us to obtain independent sequences of pseudorandom numbers  for $R$ replicas. The largest linear lattice size in our study is $L{=}70$, and about $2{\cdot}n_s {\cdot} L^2 {\approx} 10^5$ random numbers per algorithm step are used to simulate the ceiling / floor in a replica.  The total number of steps is equal to the number of energy levels, which is $2L^2{-}3 {\approx} 10^4$. The total number of random numbers per run of one replica is about $2^{30}$, which is less than the length of the Philox stream $2^{64}$.

 The total actual computation time of the Potts model ranges from 20 seconds (for $L{=}10$) and 7 minutes (for $L{=}20$) to 13 hours (for $L{=}60$).

\section{Results comparison}
\label{sec:results}

In this section, we directly compare the simulation results of the Potts model with 10 and 20 components using the WL-1/t algorithm~\cite{Wang-Landau,Bel-Per} with control of convergence~\cite{BFS} and MCPA algorithm~\cite{Rose-2019,MS-2024}. 

We will present a comparison of the energy density distribution $P(E)$, estimates of the energies in the ordered and disordered states, estimates of the critical temperature, estimates of the maximum of the specific heat, estimates of the Binder cumulant minimum, and the ratio of $P(E)$ peaks at and near the critical temperature. The estimation of the interface tension, which is inverse proportional to the correlation length, is discussed in detail.

\subsection{Specific heat and Binder cumulant scaling}

Finite-size analysis of the specific heat and the Binder cumulant was carried out in~\cite{Challa-1986} and extended in~\cite{Lee-1991},  and we use these analytical conclusions in the following sections. We estimate critical temperature from the position of the specific heat maximum and the Binder cumulant minimum, the dependence of the magnitude of the specific heat capacity maximum on the lattice size, and the magnitude of the Binder cumulant minimum. Other quantities, such as the value of the coefficients in the $L^{-2}$ corrections to the above estimates, can also be evaluated, but such a detailed analysis is beyond the scope of this paper. A corresponding analysis was presented in~\cite{Lee-1991} for the cases of the 8- and 10-component Potts model.

\subsubsection{Specific heat with q=10}

Figure~\ref{fig5} shows the specific heat in the critical region of the 10-component Potts model calculated with the WL and MCPA methods and compares it with the analytical estimate given in~\cite{Lee-1991}, expr.~(3.18). The comparison is very good, as already noted for the 10-component Potts model in the Figure 10 of the paper~\cite{Lee-1991}.

\begin{table}[]
\caption{Estimation of the maximum $C_{max}$ of specific heat and its position $T_{Cmax}$ for $q=10$. }
\begin{tabular}{|l||l|l||l|l|}
\hline
L & \multicolumn{2}{c||}{$T_{C_{max}}$} & \multicolumn{2}{c|}{$C_{max}$} \\ \hline
L  & WL & MCPA & WL & MCPA \\ \hline
16 & 0.70695(4)          & 0.7070(2)           & 73.7(1)              & 74(1)  \\ \hline
30 & 0.70298(3)          & 0.7030(2)           & 233.1(3)             & 234(2)    \\ \hline
40 & 0.70223(1)          & 0.70225(8)          & 406.1(8)            & 405(3)     \\ \hline
50 & 0.70187(2)         & 0.70188(7)          & 628(1)           & 629(5)     \\ \hline
60 & 0.70168(1)         & 0.7016(3)           & 900(1)            & 900(9) \\ \hline 
\hline
$\infty$ & 0.70124(2) & 0.70124(15) &  &   \\  \hline
exact & \multicolumn{2}{c||}{0.70123\ldots} & &\\  \hline
\end{tabular}
\label{table1} 
\end{table}

\begin{figure}[h!]
    \centering
    \includegraphics[scale=0.25]{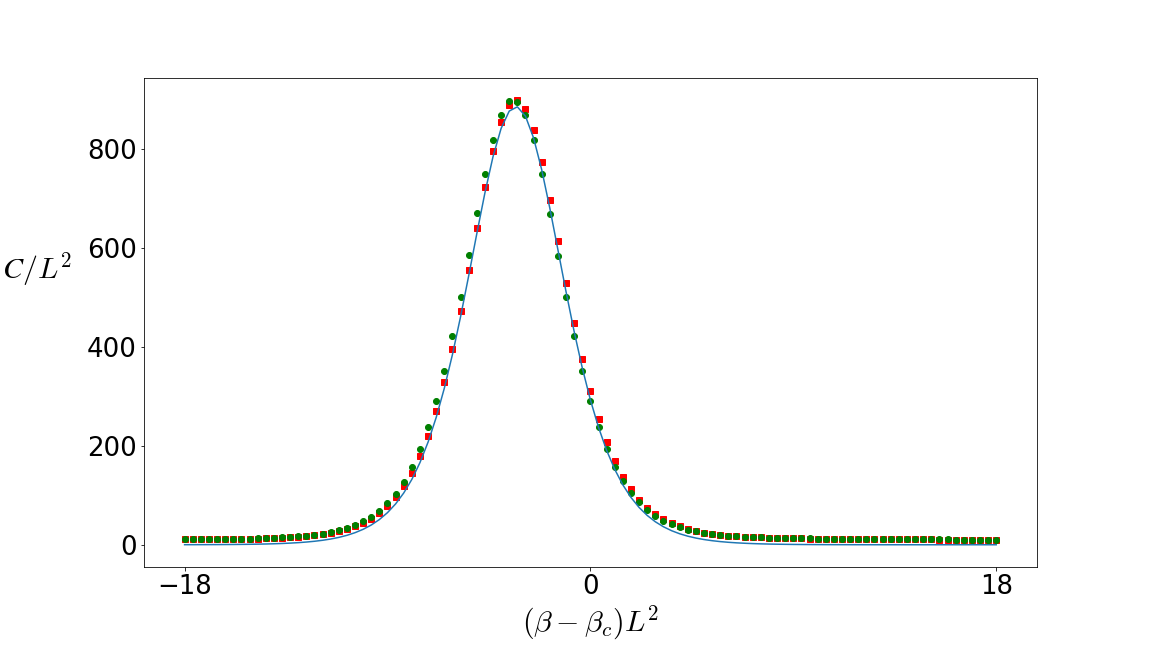}
    \caption{Specific heat as a function of inverse temperature $\beta$ in the critical region of the 10-component Potts model near the critical value $\beta_c$. Red squares are WL simulations, green circles are MCPA simulations, and the solid line is the analytical expression~(3.18) of paper~\cite{Lee-1991}. The linear size of the lattice is $L{=}60$.}
    \label{fig5}
\end{figure}
 
\begin{figure}[h!]
    \centering
    \includegraphics[scale=.3]{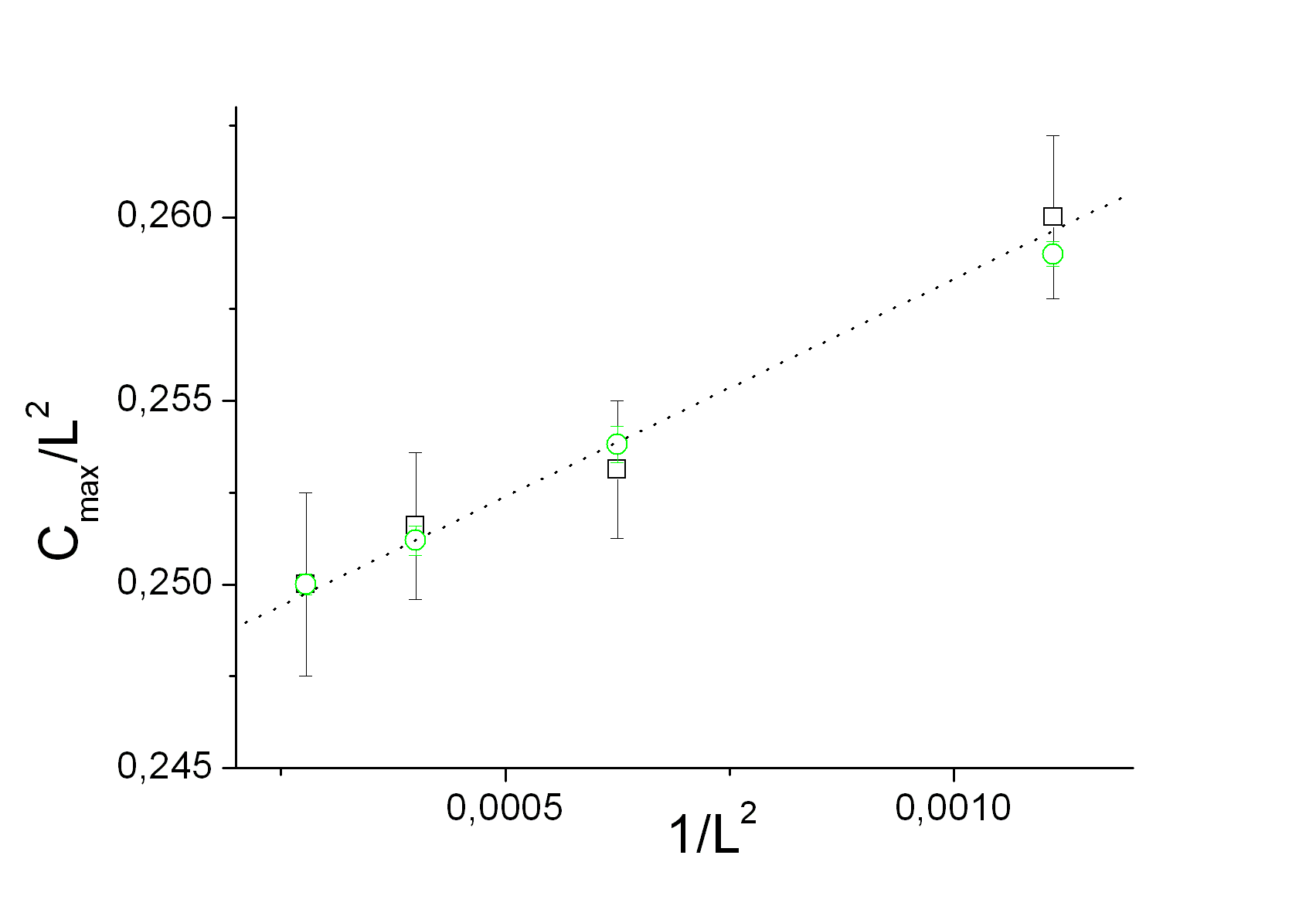}
    \caption{The ratio $C_{max}/L^2$ shown in Expr.~(\ref{eq:Cmax}) as a function of $1/L^2$: green circles correspond to $q{=}10$, black squares to $q{=}20$, the dashed line is the result of a linear fit.}
    \label{fig6}
\end{figure}

\begin{figure}[h!]
    \centering
    \includegraphics[scale=.3]{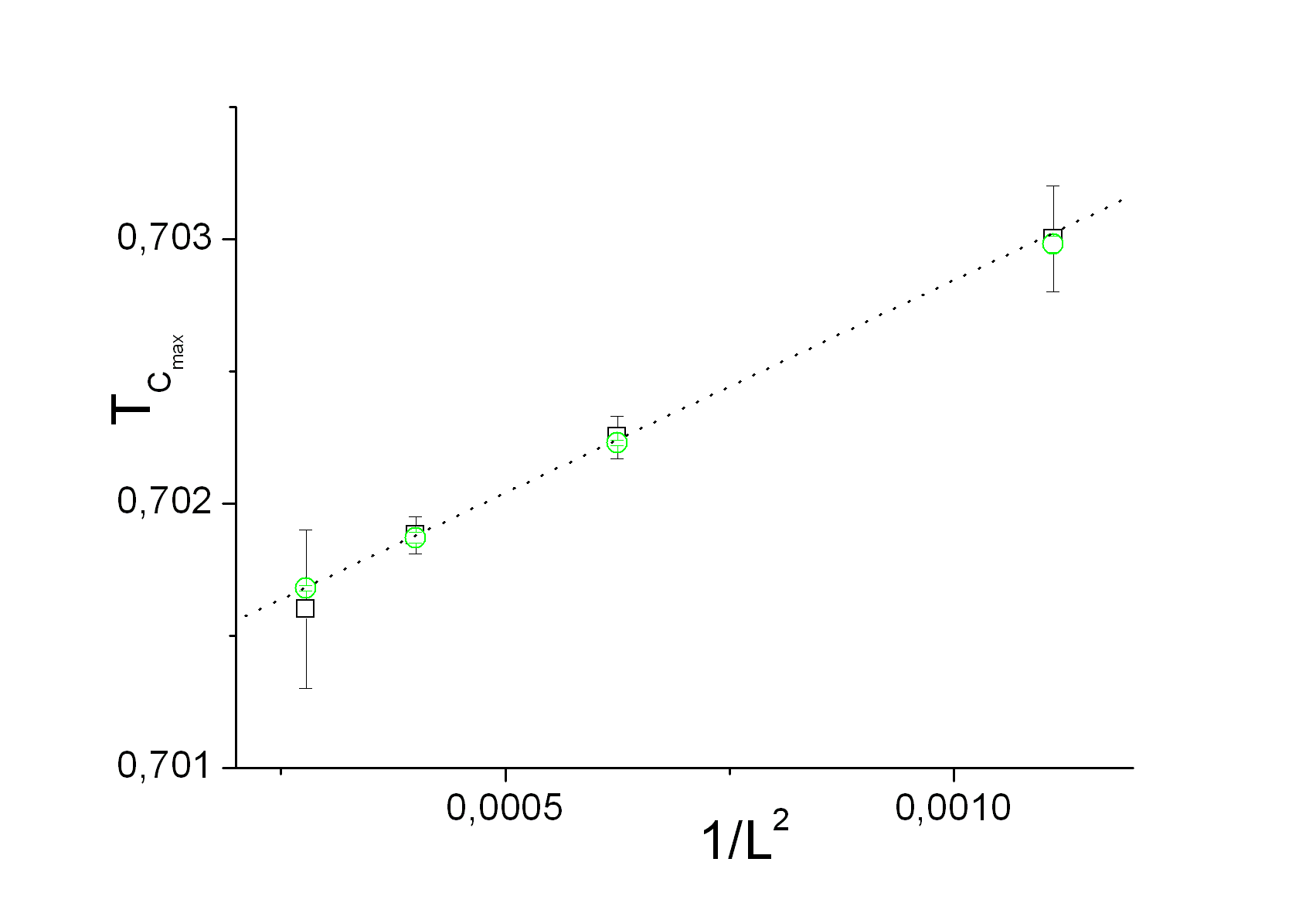}
    \caption{Estimate of the critical temperature: green circles correspond to $q{=}10$, black squares to $q{=}20$, the dashed line is the result of a linear fit.}
    \label{fig7}
\end{figure}

The maximum value of the specific heat $C_{max}$ of the Potts model in the case of a first-order phase transition depends on the volume of the system as~\cite{Challa-1986,Lee-1991} 
\begin{equation}
C_{max} /L^2\propto  \frac{(e_d-e_o)^2}{4T_c^2} = \alpha_{exact}.
\label{eq:Cmax}
\end{equation}

 The data in the fourth and fifth columns of Table~\ref{table1}  are shown in the figure~\ref{fig6} along with the error bars. The fit  excluding the smallest lattice size $L=16$ gives an estimate of $C_{max}$ with slope $\alpha{\approx}0.2470(3)$ for WL and for MCPA with slope $\alpha{\approx}0.2465(24)$, taking into account the correction terms in the fit, Expr.~(A23) from the paper~\cite{Lee-1991}. Our estimates agree well with the numerical estimate from an earlier paper~\cite{Challa-1986}, $\alpha{\approx}0.250$, and are closer to the exact value $\alpha_{exact} {=} 0.246355\ldots$.

 The critical temperature $T_c$ can be estimated from the position $T_{C_{max}}$ of the maximum of the specific heat, and the data in the second and third columns of the table~\ref{table1} are shown in the figure~\ref{fig7} along with the error bars. 
The fit results in the estimates given in the penultimate row of the table. The last row gives the exact known temperature. The estimates match the exact value to within four digits. Note that the estimates for the data obtained by the WL and MCPA methods are equally good, with the MCPA error bars being an order of magnitude larger.

\subsubsection{Specific heat with q=20}

Figure~\ref{fig8} shows the specific heat in the critical region of the 20-component Potts model calculated with the WL and MCPA methods and compares it with the analytical estimate given in~\cite{Lee-1991}, expr.~(4.5). 

\begin{figure}[h!]
    \centering
    \includegraphics[scale=0.24]{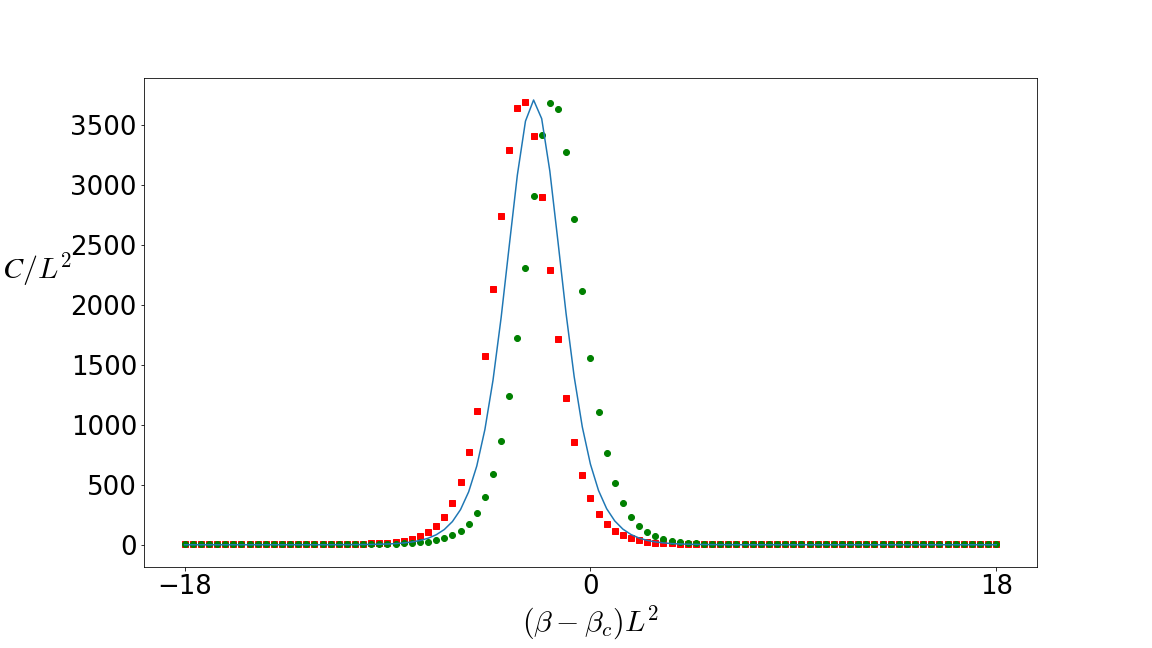}
    \caption{Specific heat as a function of inverse temperature $\beta$ in the critical region of the 20-component Potts model near the critical value $\beta_c$. Red squares are WL simulations, green circles are MCPA simulations, and the solid line is the analytical expression~(3.18) of paper~\cite{Lee-1991}. The linear size of the lattice is $L{=}60$.}
    \label{fig8}
\end{figure}

Fitting the data in the fourth and fifth columns of table~\ref{table2} yields a scaling estimate of $C_{max}$ with a slope of $\alpha{\approx}1.0306(6)$ for WL and for MCPA with a slope of $\alpha{\approx}1.0299(4)$. Our estimates agree well with the analytical prediction~\cite{Challa-1986,Lee-1991} $\alpha{=}1.02987966\ldots$ calculated from Expr.~(\ref{eq:Cmax}) using exactly known values of the critical temperature and energies of the ordered and disordered phases for the 20 components Potts model, expressions~(\ref{eq:t_crit}-\ref{eq:e_delta}).

\begin{table}[]
\caption{Estimation of the maximum $C_{max}$ of specific heat and its position $T_{Cmax}$ for $q=20$.}
\begin{tabular}{|l||l|l||l|l|}
\hline
L & \multicolumn{2}{c||}{$T_{C_{max}}$} & \multicolumn{2}{c|}{$C_{max}$} \\ \hline
L  & WL & MCPA & WL & MCPA \\ \hline
30 & 0.58934 & 0.58926(7) & 930.1  & 931(1)\\ \hline
40 & 0.58891 & 0.58886(9) & 1651.6 & 1651(3)\\ \hline
50 & 0.58869 & 0.58866(8) & 2581.9 & 2577(5)\\ \hline
60 & 0.58864 & 0.58852(5) & 3712.1 & 3713(7)\\ \hline
70 & 0.58855 & 0.5885(1)  & 5049.8 & 5053(9)\\ \hline \hline
$\infty$ & 0.58837(2) & 0.58829(2) &  &   \\  \hline
exact & \multicolumn{2}{c||}{0.5883498\ldots} & &\\  \hline
\end{tabular}
\label{table2}
\end{table}

\subsubsection{Binder cumulant with q=10}

The inverse temperature value $\beta_{B_{min}}$ at the minimum of the Binder cumulant can be used to estimate critical value, which is presented in the second and third columns of Table~\ref{table3} for the WL and MCPA methods. The magnitude of the minimum value $B_{min}$ was calculated analytically in~\cite{Lee-1991} 
\begin{equation}
B_{min} =  \frac23 -\frac{(e_o/e_d-e_d/e_o)^2}{12} + O(L^{-d})
\label{eq:Bmin}
\end{equation}
and are given in the last row of table~\ref{table3} along with the exact value of $\beta_c$.  The estimates of $\beta_{B_{min}}$ and $B_{min}$ given in the last to next row using the linear fit to the data against $1/L^2$ dependence as given in the expr.~(\ref{eq:Bmin}). The values of the inverse temperature obtained with WL  and MCPA methods agree well with the exact values. Figure~10 illustrates a linear fit to the position of the Binder cumulant minimum, which is used to estimate the value of the inverse critical temperature.  

\begin{table}[]
\caption{Comparison of $B_{min}$ and $\beta_{B_m}$ for $q=10$.}
\begin{tabular}{|l|l|l||l|l|}
\hline
L  & $\beta_{B_{min}}$,  WL & $\beta_{B_{min}}$, MCPA & $B_{min}$,     WL & $B_{min}$, MCPA \\ \hline
16 & 1.4073(8)              & 1.4072(4)     & 0.5251(2)   & 0.525(1)            \\ \hline
30 & 1.42066(3)              & 1.4207(3)    & 0.5478(2)    & 0.5474(9)           \\ \hline
40 & 1.42301(2)              & 1.4230(1)    & 0.5521(3)    & 0.5525(9)            \\ \hline
50 & 1.42411(4)              & 1.4241(1)    & 0.5542(2)    & 0.5541(9)           \\ \hline
60 & 1.4247(3)              & 1.4249(5)     & 0.5554(2)    & 0.556(1)            \\ \hline
$\infty$ & 1.42604(5) & 1.42604(24) & 0.5579(2) & 0.558(1)  \\  \hline
exact & \multicolumn{2}{c||}{1.42606\ldots}& \multicolumn{2}{c|}{0.55889\ldots}\\  \hline
\end{tabular}
\label{table3}
\end{table}

\begin{figure}[h!]
    \centering
    \includegraphics[scale=0.3]{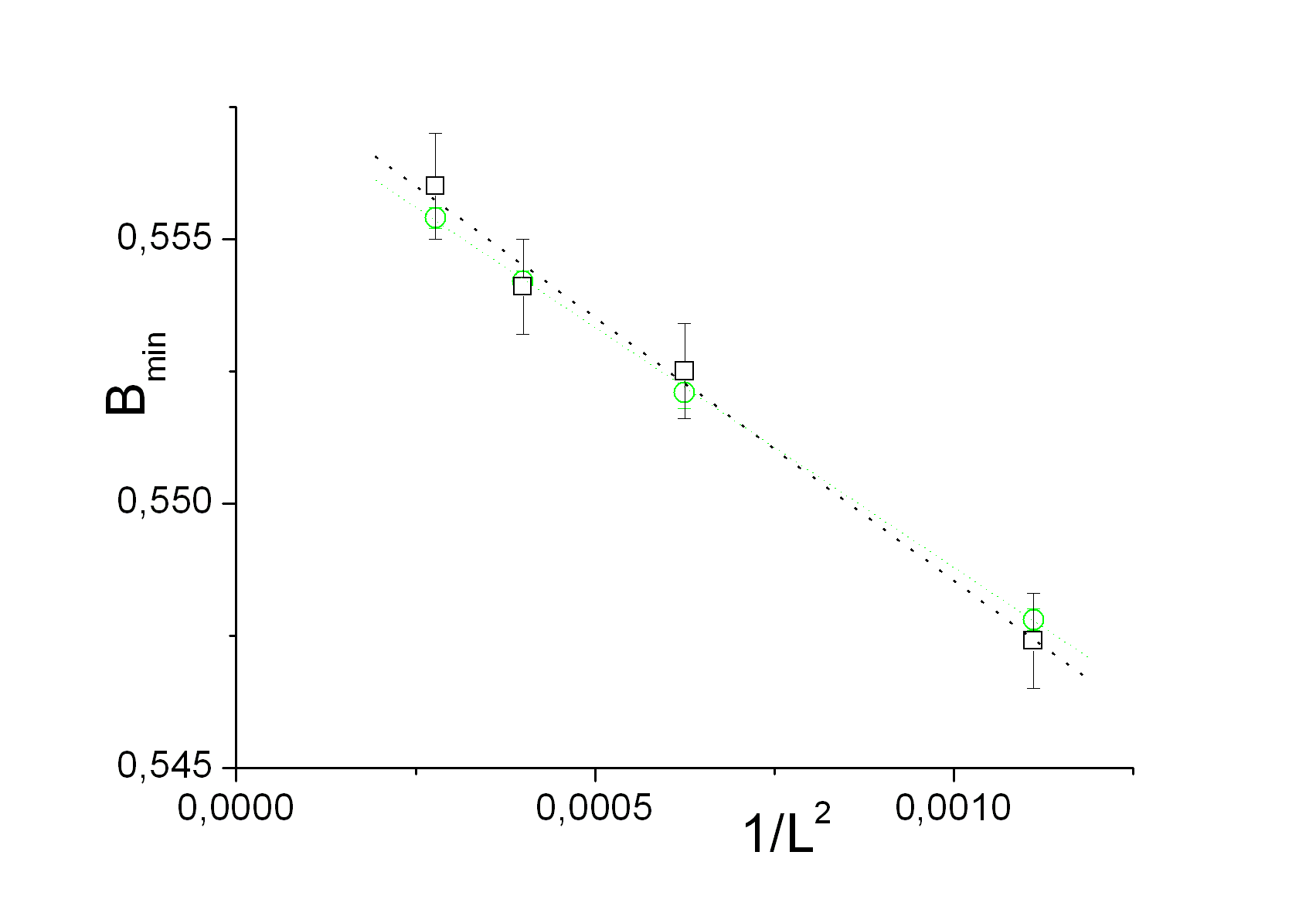}
    \caption{The Binder cumulant for $q=10$. Green circles correspond to $q{=}10$, black squares to $q{=}20$, the dashed lines is the result of a linear fits. }
    \label{fig9}
\end{figure}

\begin{figure}[h!]
    \centering
    \includegraphics[scale=0.3]{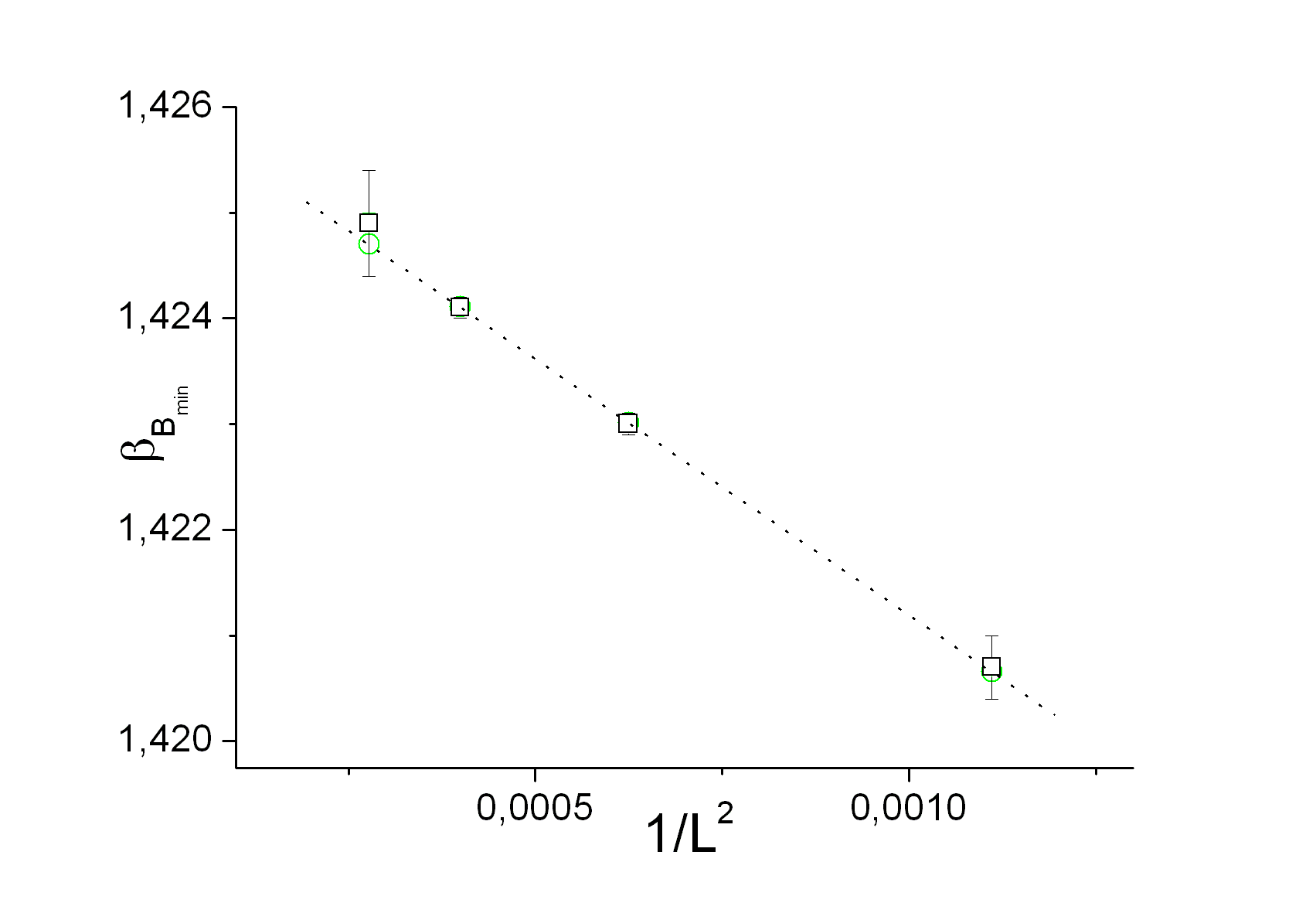}
    \caption{Estimation of the critical temperature from the position of Binder cumulant minimumn for $q=10$. Green circles correspond to $q{=}10$, black squares to $q{=}20$, the dashed line is the result of a linear fit. }
    \label{fig10}
\end{figure}

\subsubsection{Binder cumulant with q=20}

The estimates of the inverse temperature value  $\beta_{B_{min}}$ and Binder's cumulant minimum from the data in Table~\ref{table4}  agree well with the analytically known $\beta_c$ and $B_{min}$, Expr.~(\ref{eq:t_crit}) and Expr.~(\ref{eq:Bmin}) for the 20-components Potts model.

\begin{table}[]
\caption{Comparison of $B_{min}$ and $\beta_{B_m}$ for $q=20$.}
\begin{tabular}{|l|l|l||l|l|}
\hline
L  & $\beta_{B_{min}}$,  WL & $\beta_{B_{min}}$, MCPA & $B_{min}$,     WL & $B_{min}$, MCPA \\ \hline
30 & 1.694792 & 1.6950(2) & 0.1047 & 0.103(2)\\  \hline
40 & 1.696914 & 1.6971(2) & 0.1100 & 0.111(2)\\  \hline
50 & 1.697963 & 1.6981(2) & 0.1129 & 0.115(2)\\  \hline
60 & 1.698332 & 1.6987(2) & 0.1154 & 0.117(2)\\  \hline
70 & 1.698727 & 1.6990(4) & 0.1170 & 0.118(2)\\  \hline  \hline
$\infty$ & 1.6995(2) & 1.6999(2) & 0.1189(7) & 0.1209(1)  \\  \hline
exact & \multicolumn{2}{c||}{1.699669\ldots}& \multicolumn{2}{c|}{0.1197\ldots}\\  \hline
\end{tabular}
\label{table4}
\end{table}

\subsection{Probability distribution of energy}

In~\cite{Lee-1991} it is emphasized that in a finite system in the vicinity of a phase transition in the temperature range of order one, $L^d(T{-}T_C)/T {\approx} O(1)$, all states contribute significantly to the energy distribution. Therefore, the contribution of $q$ states to the ordered phase will give a peak in the probability distribution of energy $P(E)$ about $q$ times larger than the disordered state (the state with maximal possible symmetry!). Figure~\ref{fig11} shows the probability distribution of the energy $P(E)$ for linear lattice sizes $L=30$ and 60 for the 10-component Potts model, estimated by the WL and MCPA methods. The ratio of the amplitudes of the peaks $r=P(E)^{max}_o/P(E)^{max}_d$ corresponding to the ordered and disordered phases is given in Table~\ref{tb:q=10}. In addition, we give estimates of the critical ratio $r_c$~\cite{Rose-2019}
\begin{equation}
r_c=\frac{\sum{_{E<E_c}} P(E)}{\sum_{E\ge E_c} P(E)}
\label{eq:rc}
\end{equation}
taken with the energy distribution at critical temperature $T_c$ and with the sums divided by the critical energy $E_c=(E_o+E_d)/2$.

The results in Table~\ref{tb:q=10} show that both computational methods lead to reasonable estimates of this ratio. As can be seen in Figure~\ref{fig12} and Table~\ref{tb:q=20}, this is not as good for large lattice sizes in the case of the 20-component Potts model. This is because at larger lattice sizes the distributions become narrower, and the results are exponentially sensitive~\cite{LevLandau} to small variations in the computations. The same effect can be seen in the figure~\ref{fig8} for specific heat in the case of the 20-component model, where the maximum of specific heat is very close for the WL and MCPA modeling case and the analytical approximation, and this is noticeable despite the fact that the difference is only in the fourth digit  (see Table~\ref{table2}) and  simply  because of the tiny scale of the peak.

\begin{figure}[h!]
    \centering
    \includegraphics[scale=0.25]{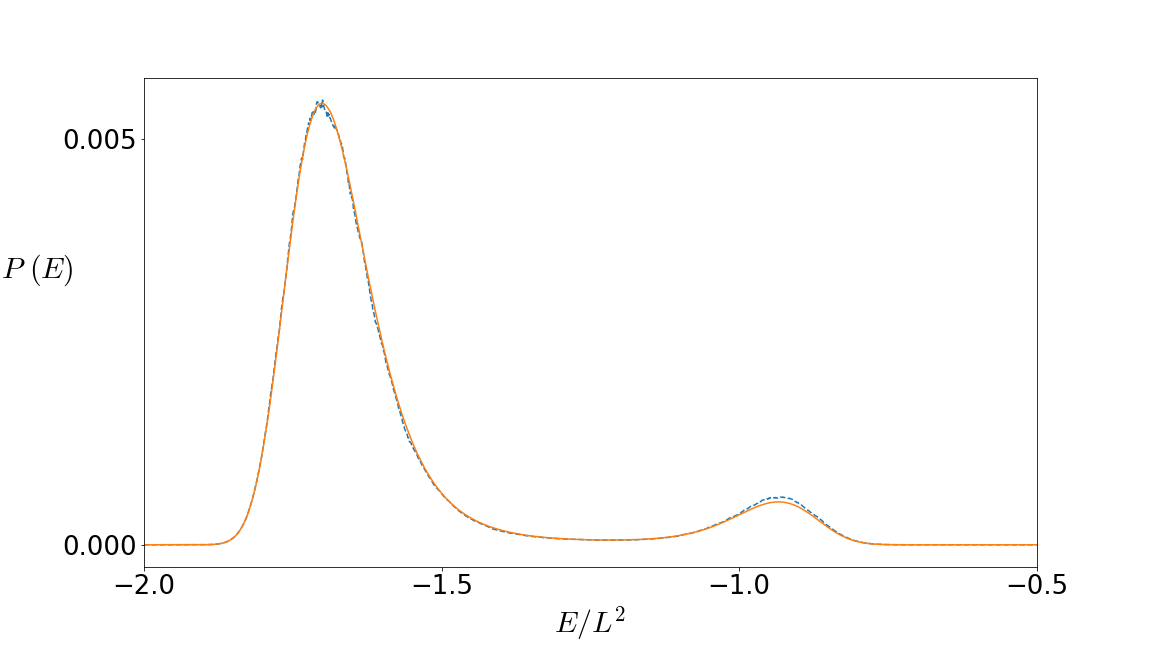} 
    \includegraphics[scale=0.25]{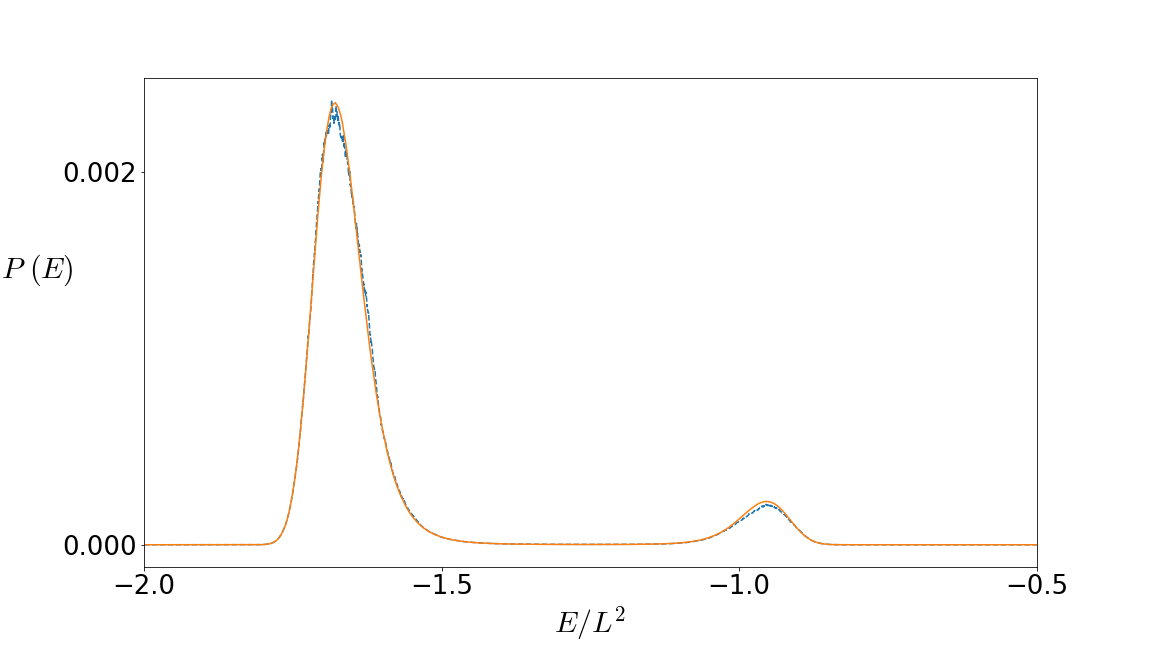} 
    \caption{Energy probability distribution $P(E)$ for 10-component Potts model for two lattice size -  top: $L=30$ and bottom: $L=60$. The solid orange line is estimation by the WL method,  the dashed blue line - by  MCPA method.}
    \label{fig11}
\end{figure}

\begin{table}[h]
    \caption{Comparison for peak ratio and $r_c$ at exact critical temperature, $q=10$.}
    \label{tb:q=10}
\begin{tabular}{|l|l|l||l|l|}
    \hline
    L & Ratio, WL & Ratio, MCPA & $r_c$, WL     & $r_c$, MCPA\\ \hline
    16 & 10.5(2) & 10.6(3) & 8.0(1) & 8.1(2)\\
    30 & 10.3(4) & 10.5(7) & 9.2(3) & 9.3(6)\\
    40 & 10.2(3) & 10.7(7) & 9.6(2) & 10.0(6)\\
    50 & 10.1(6) & 10.7(9) & 9.7(6) & 10.2(8)\\
    60 & 10.2(8) & 10.0(1) & 10.0(7) & 9.5(9)\\
    \hline
    \end{tabular}
\end{table}

\begin{figure}[h!]
    \centering
    \includegraphics[scale=0.25]{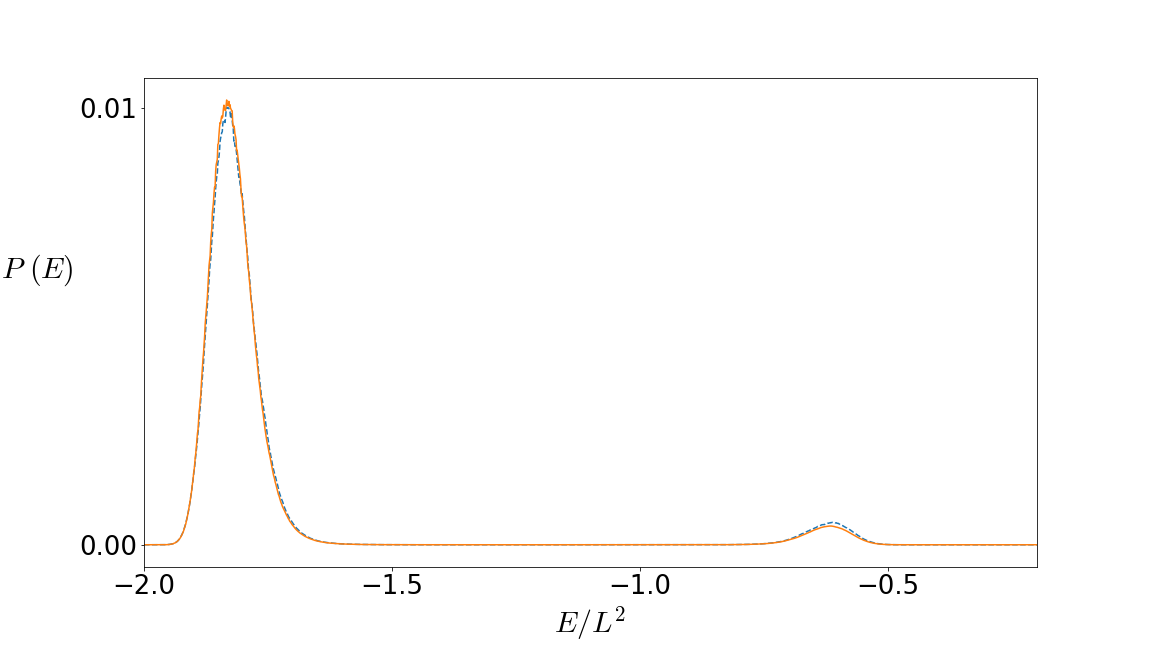} 
    \includegraphics[scale=0.25]{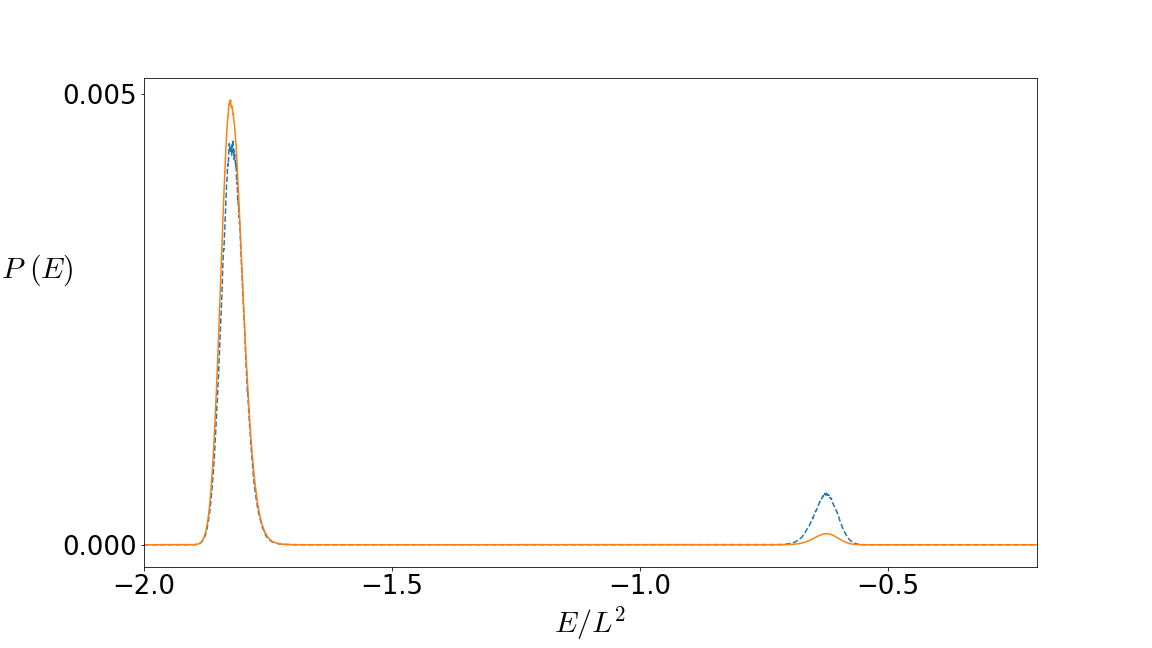} 
    \caption{Similar to Fig.~\ref{fig11} for the 20-component Potts model.}
    \label{fig12}
\end{figure}

\begin{table}[h]
    \caption{Comparison for peak ratio and $r_c$ at exact critical temperature, $q=20$.}
    \label{tb:q=20}
\begin{tabular}{|l|l|l||l|l|}
    \hline
        L & Ratio, WL & Ratio, MCPA & $r_c$, WL     & $r_c$, MCPA\\ \hline
    30 & 23.7 & 18(1) & 21.8 & 17(1)\\
40 & 24.3 & 20(3) & 22.4 & 18(3)\\
50 & 20.5 & 20(5) & 19.0 & 19(5)\\
60 & 39.9 & 12(3) & 36.5 & 11(3)\\
70 & 31.3 & 60(30) & 29.0 & 55(30)\\
    \hline
    \end{tabular}
\end{table}

\subsubsection{Estimation of energies of ordered and disordered states}

\begin{table}[]
    \caption{Estimation of $e_0$ and $e_d$ from the position of the maximum $P(e)$ calculated at the exact critical temperature,  $q=10$. These values show too little variance to pinpoint. The penultimate entry in the table is a fit to data.}
    \label{table7}
\begin{tabular}{|l|l|l||l|l|}
\hline
L  &  $E_0$, WL            & $E_0$, MCPA        & $E_d$, WL            & $E_d$, MCPA        \\ \hline
16 & -1.742(1) & -1.742(2) & -0.898(1) & -0.902(2)\\  \hline
30 & -1.700(1) & -1.700(2) & -0.934(1) & -0.934(2)\\  \hline
40 & -1.691(1) & -1.691(1) & -0.944(2) & -0.946(3)\\  \hline
50 & -1.684(1) & -1.684(1) & -0.950(1) & -0.945(1)\\  \hline
60 & -1.679(1) & -1.678(1) & -0.954(1) & -0.953(2)\\  \hline  \hline
$\infty$ & -1.659(2) & -1.656(3)& -0.974(2) & -0.975(3)   \\  \hline
exact & \multicolumn{2}{c||}{ -1.664\ldots} & \multicolumn{2}{c|}{-0.968\ldots} \\ \hline
\end{tabular}
\end{table}

\begin{table}[]
    \caption{As in the table~\ref{table7} for the 20-component Potts model.}
    \label{table8}
\begin{tabular}{|l|l|l||l|l|}
\hline
L  &  $e_0$, WL            & $e_0$, MCPA        & $e_d$, WL            & $e_d$, MCPA       
 \\ \hline
30 & -1.8333 & -1.8319(9) & -0.6156 & -0.616(1)\\  \hline
40 & -1.828  & -1.828(1)  & -0.621  & -0.620(1)\\  \hline
50 & -1.8244 & -1.8270(9) & -0.6220 & -0.6208(8)\\  \hline
60 & -1.8236 & -1.8235(7) & -0.6239 & -0.6239(8)\\  \hline
70 & -1.8225 & -1.8229(6) & -0.6241 & -0.625(1)\\  \hline  \hline
$\infty$ & -1.815(1) & -1.816(1)& -0.632(1) & -0.631(1)   \\  \hline
exact & \multicolumn{2}{c||}{ -1.82068\ldots} & \multicolumn{2}{c|}{-0.62653\ldots} \\ \hline
\end{tabular}
\end{table}

An estimate of the latent heat can be obtained from the positions $e_0(L)$, $e_d(L)$ of the maximum $P(e)$ calculated at the exact critical temperature.  Tables~\ref{table7} and \ref{table8} summarize the positions of the peaks.
We find that the data fit well with ansatz of the form $e_0=e_0(L)+b/L$ with $L{\rightarrow}\infty$ and $e_d=e_d(L)+b/L$ with $L{\rightarrow}\infty$. The fitting results give fairly good estimates of $e_0$ and $e_d$ compared to the exactly known results given in the last rows of the tables.

\subsubsection{Energy barriers and surface tension}

The first-order transition strength corresponds to the barrier height and can be estimated using the bulk free energy barrier between the ordered and disordered states as~\cite{Lee-1991}  
\begin{equation}
\Delta F(L)= ln P^{max}(L)-ln P^{min}(L)
\label{eq:lnP}
\end{equation}
where $P^{max}(E,L)$ and $P^{min}(E,L)$ are the maximum and minimum of the energy distribution taken at a value of $T$ at which the maximum corresponding to the ordered phase $P^{max}_o(L)$ coincides with the maximum corresponding to the disordered phase $P^{max}_d(L)$. Figure~\ref{fig13} shows the scaled free energy barriers $\Delta F(L)/L$ as a function of the inverse liner system size $L$, evaluated models with 10 and 20 states using the WL and MCPA algorithms. The figure~\ref{fig13} can be compared with  figure~2 from the article~\cite{Lee-1991}.

The scaled barrier limits are 0.0997(7) with MCPA and 0.0993(5) with WL for 10-state model and 0.3782(50) with MCPA and 0.3811(52) with for 20-state Potts model. This values are close to the analytical values 0.0947 and 0.371 shown in the Table II of Ref.~\cite{BJ-1992} but deviates with seven standard errors for 10-state Potts model. It should be noted that the scaled free energy barrier $\Delta F(L)/L$ is related to the interface tension and is inversely proportional to correlation length~\cite{BJ-1992}.

The deviation may be due to the fact that $P^{max}(E,L)$ and $P^{min}(E,L)$ are taken at the temperature at which $P^{max}_o(L)$ and $P^{max}_d(L)$ are equal, as suggested in Ref.~\cite{Lee-1991}. In contrast, the analytical results are calculated using data at the critical temperature $T_c$.  Therefore, we estimated barrier heights in two different ways using expression~(\ref{eq:lnP}) taking as the  $P^{max}(L)$ the  $P^{max}_o(L)$ or $P^{max}_d(L)$. Table~\ref{tab:bar-10} shows three ways of estimating barrier heights.
It seems that the estimates with heights taken at the critical temperature $T_c$ are in better agreement with the analytical results, but the scatter of the data is inconclusive. Thus, the deviations can be explained by the exponential sensitivity of barrier heights to small changes in temperature~\cite{LevLandau}. The same effect can be clearly seen in Figures~\ref{fig11} and  \ref{fig12}, which show the energy probability distributions calculated from DoS data accumulated with two different algorithms. At the same time, the results for thermodynamic quantities compare well.

\begin{table}[]
    \caption{Estimates of barrier heights for the 10-state Potts model. See text for explanation and discussion.}
    \label{tab:bar-10}
\begin{tabular}{|l||l|l|l|l|}
\hline
 algorithm &   equal heights & $P^{max}_o(L)$, $T_c$ &  $P^{max}_d(L)$, $T_c$  & analitycal \\ \hline \hline
 WL & 0.0993(5) & 0.0938(19) &   0.0941(5) & 0.0947 \\  \hline
 MCPA &  0.0997(7) & 0.0957(28)   & 0.0943(16) & 0.0947 \\  \hline
\end{tabular}
\end{table}

\begin{figure}[h!]
    \centering
    \includegraphics[scale=0.3]{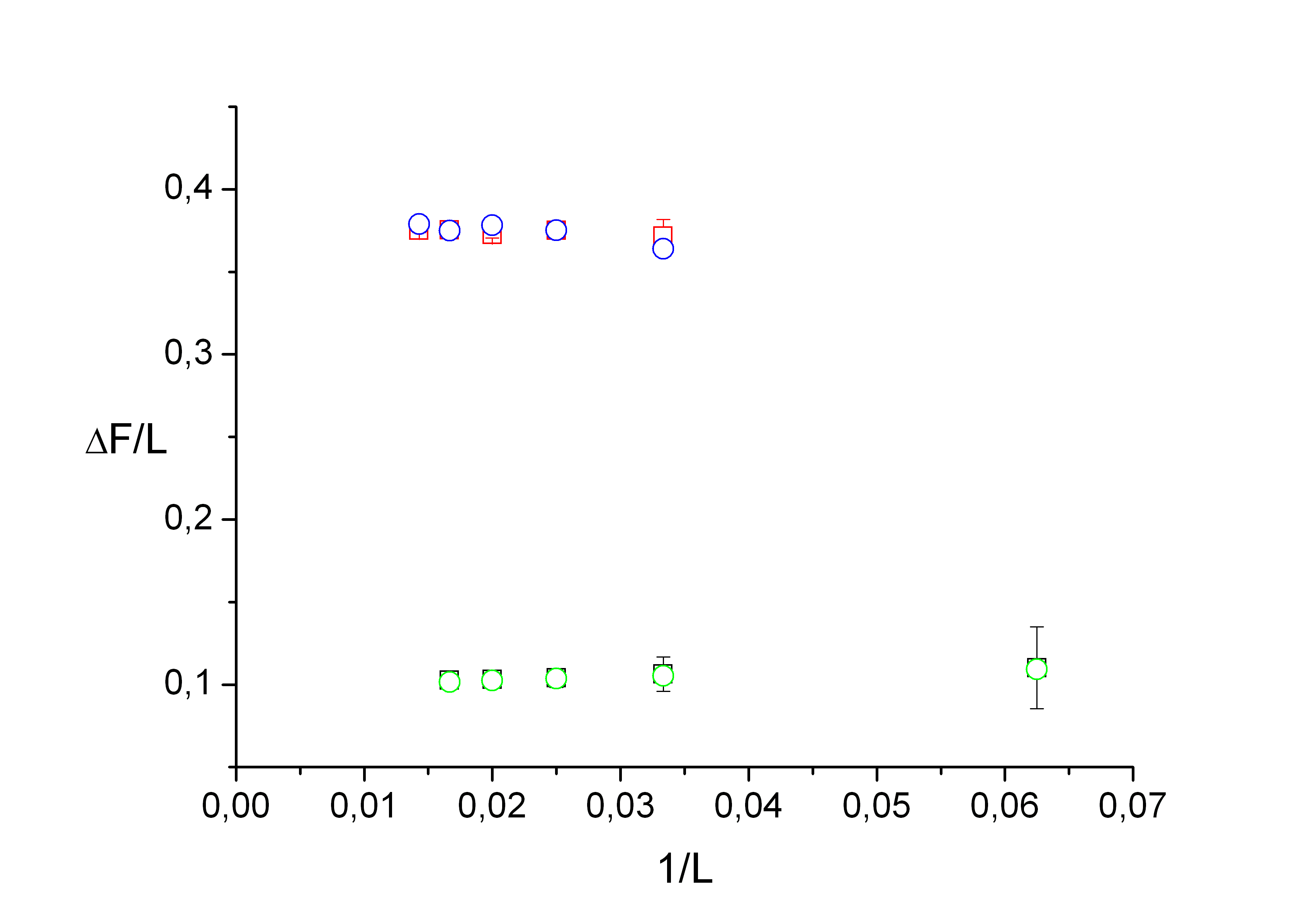} 
    \caption{The scaled free energy barriers between disordered and ordered phases for 10-state (lower symbols) and 20-state (upper symbols) Potts model. Open circles corresponds to the  estimate using data calculated with WL algorithm and open boxes - with MCPA algorithm.}
    \label{fig13}
\end{figure}

\section{Discussion}

In this paper, two methods for direct estimation of the density of states (DOS) in the spin model of statistical mechanics with discrete spectrum are comparatively analyzed. First, these are computational results for the modified Wang-Landau~\cite{Bel-Per} algorithm, a 1/t-WL algorithm, with accuracy control~\cite{BFS}. Second, these are computational results for the microcanonical population annealing algorithm~\cite{Rose-2019,MS-2024}, the MCPA algorithm. MCPA is a new algorithm and not much is known about its properties. Therefore, a direct comparison of the two algorithms is of some interest.

We stop simulation with 1/t-WL after the parameter $f$ reaches the value $f_{fin} {=} 1.00000001$, as proposed in the original version of WL algorithm~\cite{Wang-Landau}. We control the convergence of DoS by computing the largest eigenvalue of the transfer matrix in the energy spectrum~\cite{BFS} (see Figures ~\ref{fig1} and \ref{fig2}). The number of spin update operations ranges from $5{\cdot} 10^{10}$ to $72{\cdot} 10^{10}$ for lattice sizes from $L{=}16$ to $L{=}60$.

The MCPA simulation is completed using a ceiling/floor procedure, reaching the lowest/highest energy level. The number of operations in this case is not random. The number of spin updates in each replica is equal to the product of the number of spins $L^2$, the number of MCMC iterations $n_s$, and the number of energy levels $N_E{=}2L^2{-}3$ for models with 10 and 20 components. For a typical number of iterations $n_s{=}10$, the number of spin flips per replica varies from $16{\cdot} 10^{6}$ to $480{\cdot} 10^{6}$, and the linear size of the system varies from $L{=}30$ to $L{=}70$. The total number of spin updates must be multiplied by the number of replicas $R{=}2^{17}{\approx} 1.3{\cdot}10^5$. Thus, the total number of spin updates is comparable with the WL updates mentioned above to an order-of-magnitude precision. 

The estimates of thermodynamic quantities computed with the WL and MCPA algorithms have the same accuracy, as shown in the paper, and the advantage of MCPA is that it requires fewer operations per thread than the WL algorithm. Of course, the threads in both cases are different: WL uses CPU threads, while MCPA uses GPU threads. Therefore, it is neither an easy way to compare the computational effectiveness of algorithms. 
 The typical time per computational run for DoS estimation of a 10-state Potts model on the $L{=}16$ lattice is 5942 seconds with the WL algorithm and 163 seconds with the MCPA algorithm, with a real-time speedup of about 32. A similar run on the $L{=60}$ lattice is 413312 s with the WL algorithm and 47091 s with the MCPA algorithm, with a real-time acceleration of about 9. The decrease in speedup with increasing lattice size is due to more memory swapping between on-board and GPU memory.

In addition, MCPA can be used to estimate magnetization and its moments, i.e. magnetic susceptibility and magnetic Binder cumulant~\cite{Binder-C} from a single run and averaging as functions of the replica pool for energy $E'$ (the culling pool), yielding $M(E')$, $\xi(E')$, and $B(E')$. In contrast, to extract the magnetic sector of thermodynamic variables using the WL algorithm, the two-dimensional distribution function $g(E,M)$~\cite{gEM} must be evaluated for both energy and magnetization, which significantly increases the required computation time.  This is another advantage of MCPA over WL.

\begin{acknowledgments}
The research was supported by the Russian Science Foundation grant 22-11-00259. 

The simulation was carried out using high-performance computing  resources of the National Research University Higher School of Economics. 
\end{acknowledgments}


\begin{thebibliography}{99}
\bibitem{LevLandau} In some cases, the least significant digits can be important. According to the memories of colleagues, Lev Landau used to say that "money remains in an exponential function'', so counting it to the last coin is exponentially important.
\end{thebibliography}
\end{document}